\documentclass[amsmath,pra,twocolumn,showpacs]{revtex4}
\usepackage{graphics}

\begin{document}     

\title{Phonon-mediated decay of an atom in a surface-induced potential}

\author{Fam Le Kien,$^{1,*}$ S. Dutta Gupta,$^{1,2}$ and K. Hakuta$^{1}$} 
\affiliation{
$^1$Department of Applied Physics and Chemistry, 
University of Electro-Communications, Chofu, Tokyo 182-8585, Japan\\
$^2$School of Physics, University of Hyderabad, Hyderabad, India}

\date{\today}

\begin{abstract}
We study phonon-mediated transitions between translational levels of an atom in a surface-induced potential. We present a general master equation governing the dynamics of the translational states of the atom.
In the framework of the Debye model, we derive
compact expressions for the rates for both upward and downward transitions. 
Numerical calculations for the transition rates are
performed for a deep silica-induced potential allowing for a large number of bound levels as well as free states of a cesium atom. The total absorption rate is shown to be determined mainly by the bound-to-bound
transitions for deep bound levels and by bound-to-free transitions for shallow bound
levels. Moreover, the phonon emission and absorption processes can be
orders of magnitude larger for deep bound levels as compared to the shallow bound
ones. We also study various types of transitions from free states. 
We show that, for thermal atomic cesium with temperature in the range from 100 $\mu$K 
to 400 $\mu$K in the vicinity of a silica surface with temperature of 300 K, 
the adsorption (free-to-bound decay) rate is about two times larger than the  heating (free-to-free upward decay) rate, while the cooling (free-to-free downward decay) rate is negligible.

\end{abstract}

\maketitle

\section{Introduction}

Over the past few years, tight confinement of cold atoms has drawn considerable attention. The interest in this area is motivated not only by the fundamental nature of the problem, but also by its potential applications in atom optics and quantum information. A  method for microscopic trapping and guiding of individual atoms along a nanofiber has been proposed \cite{our traps}. Surface--atom quantum electrodynamic effects have constituted another interesting area, where a great deal of work has been carried out. Modification of spontaneous emission 
of an atom \cite{cesium decay} and radiative exchange between two distant atoms \cite{two atoms} mediated by a nanofiber have been investigated. Surface-induced deep potentials have played a major role and have received due attention in recent years. Oria \textit{et al.} have studied various theoretical schemes to load atoms into such potentials \cite{Lima,Oria2006}. A rigorous theory of spontaneous decay of an atom in a surface-induced potential 
invoking the density-matrix formalism has been developed \cite{boundspon}. 
The role of interference between the emitted and reflected fields and also the role of transmission into the evanescent modes were identified. Further calculations on the excitation spectrum have been carried out \cite{spectrum}. Bound-to-bound transitions were shown to lead to significant effects like a large red tail of the excitation spectrum as compared to the weak consequences of free-to-bound transitions. A crucial step in this direction was the experimental observation of the excitation spectrum and the channeling of the fluorescent photons along the nanofiber \cite{Kali}, opening up avenues for novel quantum information devices.

In most of the problems involving surface--atom interaction, the macroscopic surface is usually kept at room temperature. Thus the pertinent question that can be asked is what would be the effect of heating on the cold atoms. It is understood that transfer of heat to the trapped atoms will lead to a change in the occupation probability of the vibrational levels as well as their coherence. Phonon-induced changes in the populations of the vibrational levels have been studied by several groups \cite{Oria2006,Henkel,Gortel}. In a nice and compact treatment based on the dyadic Green function and the Fermi golden rule, Henkel \textit{et al.} showed that the effects can be very different depending on the nature of the atomic/molecular species \cite{Henkel}. The time scales for various species were estimated. It should be stressed that the trap considered by Henkel \textit{et al.} was not necessarily a surface trap and misses out on many of the aspects of the surface--atom interaction \cite{Henkel}. Based on the assumption that the surface--atom interaction can be represented by a Morse potential, the phonon-mediated decay was estimated by Oria \textit{et al.} \cite{Oria2006}. Their estimate was based on the formalism developed by Gortel \textit{et al.} \cite{Gortel}. However, all the previous theories focus on only the transition rates and thus are not general enough. In this paper, we present a general density-matrix formalism to calculate the phonon-mediated decay of populations as well as the changes in coherence. We derive the relevant master equation for the density matrix of the atom. We emphasize that our density-matrix equation describes the full dynamics of the coupling between trapped atoms and phonons and does not assume any particular form of the trapping potential. Under the Debye approximation, we derive compact expressions for the phonon-mediated decay rates. Numerical calculations are carried out assuming the potential model considered in \cite{Lima}. In contrast to the previous work, we include a large number of vibrational levels due to the deep surface--atom potential. We show that there can be significant differences in the decay rates when the initial level is chosen as one of the shallow or deep bound levels.
We also calculate and analyze the decay rates for various types of transitions from free states.

The paper is organized as follows. In Sec.\ \ref{sec:model} we describe the model.
In Sec.\ \ref{sec:analytical} we derive the basic dynamical equations for the  phonon-mediated decay processes. 
In Sec.\ \ref{sec:numerical} we present the results of numerical calculations. 
Our conclusions are given in Sec.~\ref{sec:summary}.

\section{Description of the model system}
\label{sec:model}

We assume the whole space to be divided into two regions, namely,
the half-space $x<0$, occupied by a nondispersive nonabsorbing dielectric medium (medium 1), and the half-space $x>0$,  occupied by vacuum (medium 2).
We examine a single atom moving in the empty half-space $x>0$. We assume that the atom is in a fixed internal state $|i\rangle$ with energy $\hbar\omega_i$.
Without loss of generality, we assume 
that the energy of the internal state $|i\rangle$ is zero, i.e. $\omega_i=0$. 
We describe the interaction between the atom and the surface.  
We first consider the surface-induced interaction potential and then add the atom-phonon interaction.

\subsection{Surface-induced interaction potential}
\label{sec:model A}

In this subsection, we describe the interaction between the atom and the surface in the case where thermal vibrations of the surface are absent.
The potential energy of the surface--atom interaction is
a combination of a long-range van der Waals attraction 
and a short-range repulsion \cite{Hoinkes}. 
Despite a large volume of research on the surface--atom interaction, 
due to the complexity of surface physics and the lack of data, the actual form of the potential is yet to be ascertained \cite{Hoinkes}. For the purpose of numerical demonstration of our formalism, we choose the following model for the potential \cite{Lima,Hoinkes}:
\begin{equation}
U(x)=A e^{-\alpha x}-\frac{C_{3}}{x^3}.
\label{h1}
\end{equation} 
Here, $C_3$ is the van der Waals coefficient, while $A$ and $\alpha$ determine the height and range, respectively, of the surface repulsion. 
The potential parameters $C_{3}$, $A$, and $\alpha$ depend 
on the nature of the dielectric and the atom. 
In numerical calculations, we use the parameters of fused silica, for the dielectric,
and the parameters of ground-state atomic cesium, for the atom.
The parameters for the interaction between silica and ground-state atomic cesium are theoretically estimated to be $C_{3}=1.56$ kHz $\mu$m$^3$, $A=1.6\times 10^{18}$ Hz, 
and $\alpha=53$ nm$^{-1}$ \cite{boundspon}. 
 
We introduce the notation $\varphi_{\nu}(x)$ for the  eigenfunctions of the center-of-mass motion of the atom in the potential $U(x)$. 
They are determined by the stationary Schr\"{o}dinger equation
\begin{equation}
\left[-\frac{\hbar^2}{2m}\frac{d^2}{dx^2}+U(x)\right]\varphi_{\nu}(x)
=\mathcal{E}_{\nu}\varphi_{\nu}(x).
\label{h2}
\end{equation}
Here $m$ is the mass of the atom. 
In the numerical example with atomic cesium, we have $m=132.9$ a.u. $=2.21\times 10^{-25}$ kg. The eigenvalues $\mathcal{E}_{\nu}$ are the center-of-mass energies
of the translational levels of the atom. 
These eigenvalues are the shifts of the energies of the translational 
levels from the energy of the internal state $|i\rangle$.
Without loss of generality, we assume that
the center-of-mass eigenfunctions $\varphi_{\nu}(x)$ are real functions, i.e. 
$\varphi_{\nu}^*(x)=\varphi_{\nu}(x)$.

In Fig.~\ref{fig1}, we show the potential $U(x)$ and the wave functions 
$\varphi_{\nu}(x)$ of a number of bound levels  with energies 
in the range from $-1$ GHz to $-5$ MHz. 
We also plot the wave function of a free state with energy of about 4.25 MHz. 
In order to have some estimate about the spatial extent 
of a wave function $\varphi_{\nu}(x)$, we define a crossing point $x_{\mathrm{cross}}$, 
which corresponds to the rightmost solution of the equation $U(x)=\mathcal{E}_\nu$.
Note that, for shallow levels, the wave function generally peaks close to the point
$x_{\mathrm{cross}}$. 
We plot the eigenvalue modulus $|\mathcal{E}_{\nu}|$ and the crossing point 
$x_{\mathrm{cross}}$ in Figs.~\ref{fig2}(a)
and \ref{fig2}(b), respectively. It is clear from the figure that,
for $\nu$ in the range from 0 to 300, the eigenvalue  varies dramatically from about
158 THz to about 322 kHz, while the wave function extends only up to 170 nm.

\begin{figure}[tbh]
\begin{center}
 \end{center}
\caption{Energies and wave functions of the center-of-mass motion of an atom in a surface-induced potential. The parameters of the potential are 
$C_{3}=1.56$ kHz $\mu$m$^3$, $A=1.6\times 10^{18}$ Hz, and $\alpha=53$ nm$^{-1}$.
The mass of the atom is $m=2.21\times 10^{-25}$ kg. 
We plot bound levels  with energies 
in the range from $-1$ GHz to $-5$ MHz and also a free state with energy of about 4.25 MHz.
}
\label{fig1}
\end{figure}

\begin{figure}[tbh]
\begin{center}
 \end{center}
\caption{Eigenvalue modulus $|\mathcal{E}_{\nu}|$ (a) and crossing point 
$x_{\mathrm{cross}}$ (b) as functions of the vibrational quantum number $\nu$.
The parameters used are as in Fig.~\ref{fig1}.
}
\label{fig2}
\end{figure}

We introduce the notation $|\nu\rangle=|\varphi_{\nu}\rangle$
and $\omega_{\nu}=\mathcal{E}_{\nu}/\hbar$ for the state vectors and frequencies of
translational levels. Then, the Hamiltonian of the atom in the surface-induced potential 
can be represented in the diagonal form
\begin{equation}
H_A=\sum_{\nu}\hbar\omega_{\nu}\sigma_{\nu\nu}.
\label{h3}
\end{equation}
Here, $\sigma_{\nu\nu}=|\nu\rangle\langle \nu|$ is the population operator for the translational level $\nu$.
We emphasize that the summation over $\nu$  includes both the discrete 
($\mathcal{E}_{\nu}<0$) and continuous ($\mathcal{E}_{\nu}>0$) spectra. 
The levels $\nu$ with $\mathcal{E}_{\nu}<0$ 
are called the bound (or vibrational) levels. 
In such a state, the atom is bound to the surface. 
It is vibrating, or more exactly, moving back and forth
between the walls formed by the van der Waals part and the repulsive part of the potential.
The levels $\nu$ with $\mathcal{E}_{\nu}>0$ 
are called the free (or continuum) levels.
The center-of-mass wave functions of the bound states are normalized to unity.
The center-of-mass wave functions of the free states are normalized to the delta function of energy.

\subsection{Atom--phonon interaction}
\label{sec:model B}

In this subsection, we incorporate the thermal vibrations of the solid into the model. Due to the thermal effects, the surface of the dielectric vibrates. The surface-induced potential for the atom is then $U(x-x_s)$, where $x_s$ is the displacement of the surface from the mean position $\langle x_s\rangle=0$. 
We approximate the vibrating potential $U(x-x_s)$ 
by expanding it to the first order in $x_s$, 
\begin{equation}
U(x-x_s)=U(x)-U'(x)x_s.
\label{h4}
\end{equation}
The first term, $U(x)$, when combined with the kinetic energy $p^2/2m$, yields
the Hamiltonian $H_A$ [see Eq. (\ref{h3})], 
which leads to the formation of translational levels of the atom.
The second term, $-U'(x)x_s$,
accounts for the thermal effects in the interaction of the atom with the solid. 
Note that the quantity $F=-U'(x)$ is the force of the surface upon the atom. Hence, the force of the atom upon the surface
is $-F=U'(x)$ and, consequently, $U'(x)x_s$ is the work required to displace the surface for a small distance $x_s$.

It is well known that, for a smooth surface, 
the gas atom interacts only with the phonons polarized along the $x$ direction \cite{Gortel}.
In the harmonic approximation, we have
\begin{equation}
x_s=\sum_{\mathbf{q}}\left(\frac{\hbar}{2MN\omega_{\mathbf{q}}}\right)^{1/2}(b_{\mathbf{q}}e^{i\mathbf{qR}}+b_{\mathbf{q}}^\dagger e^{-i\mathbf{qR}}).
\label{h5}
\end{equation} 
Here, $M$ is the mass of a particle of the solid, $N$ is the particle number density, 
$\omega_{\mathbf{q}}$ and $\mathbf{q}$ are the frequency and wave vector
of the $x$-polarized acoustic phonons, respectively, 
$\mathbf{R}=(0,y,z)$ is the lateral component of the position vector $(x,y,z)$ of the atom, and $b_{\mathbf{q}}$ and  $b_{\mathbf{q}}^\dagger$ are the annihilation and creation phonon operators, respectively.
Without loss of generality, we choose $\mathbf{R}=0$. 
Meanwhile, the operator $U'$ can be decomposed as 
$U'=\sum_{\nu\nu'}\sigma_{\nu\nu'}\langle\nu|U'|\nu'\rangle$, where
$\sigma_{\nu\nu'}=|\nu\rangle\langle\nu'|$ is the operator for 
the translational transition $\nu\leftrightarrow\nu'$.
Hence, the energy term $-U'(x)x_s$ leads to the atom--phonon interaction Hamiltonian \cite{Gortel}
\begin{equation}
H_I=\hbar\sum_{\mathbf{q}}\frac{1}{\sqrt{\omega_{\mathbf{q}}}}S (b_{\mathbf{q}}+b_{\mathbf{q}}^\dagger),
\label{h6}
\end{equation}
with
\begin{equation}
S=\sum_{\nu\nu'}g_{\nu\nu'}\sigma_{\nu\nu'}.
\label{h11a}
\end{equation}
Here we have introduced the atom--phonon coupling coefficients 
\begin{equation}
g_{\nu\nu'}=\frac{F_{\nu\nu'}}{\sqrt{2MN\hbar}},
\label{h7}
\end{equation}
with
\begin{equation}
F_{\nu\nu'}=-\int_{-\infty}^{\infty}\varphi_{\nu}(x)U'(x)\varphi_{\nu'}(x)dx
\label{h7c}
\end{equation} 
being the matrix elements for the force of the surface upon the atom.
We note that $F_{\nu\nu'}=-m\omega_{\nu\nu'}^2x_{\nu\nu'}$,
where $x_{\nu\nu'}=\langle\nu|x|\nu'\rangle$
and $\omega_{\nu\nu'}=\omega_{\nu}-\omega_{\nu'}$
are the surface--atom dipole matrix element and the translational transition frequency, respectively.  
Hence, the coupling coefficient $g_{\nu\nu'}$
depends on the dipole matrix element $x_{\nu\nu'}$ and
the transition frequency $\omega_{\nu\nu'}$.
Since $\omega_{\nu\nu}=0$, we have $g_{\nu\nu}=0$. 
 
We note that the Hamiltonian of the $x$-polarized acoustic phonons is given by
\begin{equation}
H_B=\sum_{\mathbf{q}} \hbar\omega_{\mathbf{q}} b_{\mathbf{q}}^\dagger b_{\mathbf{q}}.
\label{h8}
\end{equation}
The total Hamiltonian of the atom--phonon system is 
\begin{equation}
H=H_A+H_I+H_B.
\label{h8a}
\end{equation}
We use the above Hamiltonian to study the phonon-mediated decay of the atom.

\section{Dynamics of the atom}
\label{sec:analytical}

In this section, we present the basic equations for the  phonon-mediated decay processes. We derive a general master equation for 
the reduced density operator of the atom in subsection \ref{sec:basic A},
obtain analytical expressions for the relaxation rates and frequency shifts
in subsection \ref{sec:basic B}, and calculate the rates and the shifts in the framework of the Debye model in subsection \ref{sec:basic C}.  

\subsection{Master equation}
\label{sec:basic A}

In the Heisenberg picture, the equation for the phonon operator $b_{\mathbf{q}}(t)$ is
\begin{equation}
\dot{b}_{\mathbf{q}}(t)=-i\omega_{\mathbf{q}}b_{\mathbf{q}}(t)-\frac{i}{\sqrt{\omega_{\mathbf{q}}}}S(t),
\label{h27}
\end{equation}
which has a solution of the form
\begin{equation}
b_{\mathbf{q}}(t)=b_{\mathbf{q}}(t_0)e^{-i\omega_{\mathbf{q}}(t-t_0)}-iW_{\mathbf{q}}(t).
\label{h28}
\end{equation}
Here, $t_0$ is the initial time
and $W_{\mathbf{q}}$ is given by
\begin{equation}
W_{\mathbf{q}}(t)=\frac{1}{\sqrt{\omega_{\mathbf{q}}}}\int_{t_0}^te^{-i\omega_{\mathbf{q}}(t-\tau)}S(\tau)\,d\tau.
\label{h29}
\end{equation}
Consider an arbitrary atomic operator $\mathcal{O} $ which acts only on the atomic states but not on the phonon states.
The time evolution of this operator is governed by the Heisenberg equation
\begin{equation}
\frac{\partial  \mathcal{O}(t)}{\partial t} = 
\frac{\mathrm{i}}{\hbar}[H_A(t)+H_I(t),\mathcal{O}(t)],
\label{h30}
\end{equation}
which, with account of Eqs. (\ref{h6}) and (\ref{h28}), yields 
\begin{eqnarray}
\lefteqn{\frac{\partial  \mathcal{O}(t)}{\partial t} = \frac{\mathrm{i}}{\hbar}[H_A(t),\mathcal{O}(t)]}
\nonumber\\&&\mbox{}
+\sum_{\mathbf{q}}\frac{i}{\sqrt{\omega_{\mathbf{q}}}} [S(t),\mathcal{O}(t)]
[b_{\mathbf{q}}(t_0)e^{-i\omega_{\mathbf{q}}(t-t_0)}-iW_{\mathbf{q}}(t)] 
\nonumber\\&&\mbox{}
-\sum_{\mathbf{q}}\frac{i}{\sqrt{\omega_{\mathbf{q}}}} 
[b_{\mathbf{q}}^\dagger(t_0)e^{i\omega_{\mathbf{q}}(t-t_0)}+iW_{\mathbf{q}}^\dagger(t)][\mathcal{O}(t),S(t)].
\nonumber\\
\label{h31}
\end{eqnarray}

We assume the initial density of the atom--phonon system to be the direct product state 
\begin{equation}
\rho_{\Sigma}(t_0)= \rho(t_0) \rho_B(t_0), 
\label{h32}
\end{equation}
with the atom in an arbitrary state $\rho(t_0)$ and the phonons in a thermal state 
\begin{equation}
\rho_B(t_0)=Z^{-1}\exp[-H_B(t_0)/k_BT].
\label{h33}
\end{equation}
Here, $Z$ is the normalization constant and $T$ is the temperature of the phonon bath.
For the initial condition (\ref{h32}), 
the Bogolubov's lemma \cite{Bogolubov}, applied to an arbitrary operator $\Theta(t)$, 
asserts the following:
\begin{equation}
\langle \Theta(t)b_{\mathbf{q}}(t_0) \rangle=\bar{n}_{\mathbf{q}}\langle [b_{\mathbf{q}}(t_0),\Theta(t)] \rangle,
\label{h34}
\end{equation}
where the mean number of phonons in the mode $\mathbf{q}$ is given by
\begin{equation}
\bar{n}_{\mathbf{q}}=\frac{1}{\exp(\hbar\omega_{\mathbf{q}}/k_BT)-1}.
\label{h22}
\end{equation}
Let $\Theta$ be an atomic operator. We then have the commutation relation $[b_{\mathbf{q}}(t),\Theta(t)]=0$, which yields
\begin{equation}
[b_{\mathbf{q}}(t_0),\Theta(t)]=ie^{i\omega_{\mathbf{q}}(t-t_0)}[W_{\mathbf{q}}(t),\Theta(t)].
\label{h35}
\end{equation}
Combining Eq. (\ref{h34}) with Eq. (\ref{h35}) leads to
\begin{equation}
\langle \Theta(t)b_{\mathbf{q}}(t_0) \rangle =
ie^{i\omega_{\mathbf{q}}(t-t_0)}\bar{n}_{\mathbf{q}}\langle [W_{\mathbf{q}}(t),\Theta(t)] \rangle.
\label{h36}
\end{equation}
We perform the quantum mechanical averaging for expression (\ref{h31}) 
and use Eq. (\ref{h36}) to eliminate the phonon operators $b_{\mathbf{q}}(t_0)$ and $b_{\mathbf{q}}^\dagger(t_0)$. 
The resulting equation can be written as
\begin{eqnarray}
\lefteqn{\frac{\partial  \langle\mathcal{O}(t)\rangle}{\partial t} 
= \frac{\mathrm{i}}{\hbar}\langle[H_A(t),\mathcal{O}(t)]\rangle}
\nonumber\\&&\mbox{}
+\sum_{\mathbf{q}}\frac{\bar{n}_{\mathbf{q}}+1}{\sqrt{\omega_{\mathbf{q}}}} 
\langle[S(t),\mathcal{O}(t)]W_{\mathbf{q}}(t)+W_{\mathbf{q}}^\dagger(t)[\mathcal{O}(t),S(t)] \rangle
\nonumber\\&&\mbox{}
+\sum_{\mathbf{q}}\frac{\bar{n}_{\mathbf{q}}}{\sqrt{\omega_{\mathbf{q}}}} 
\langle W_{\mathbf{q}}(t) [\mathcal{O}(t),S(t)] + [S(t),\mathcal{O}(t)]W_{\mathbf{q}}^\dagger(t)\rangle.
\nonumber\\
\label{h37}
\end{eqnarray}
We note that Eq. (\ref{h37}) is exact. 
It does not contain phonon operators explicitly.
The dependence on the phonon operators is hidden in the time
shift of the operator $S(\tau)$ in expression (\ref{h29}) 
for the operator $W_{\mathbf{q}}(t)$. 

We now show how the dependence of the operator $W_{\mathbf{q}}(t)$ on the phonon operators  can be approximately eliminated. 
We assume that the atom--phonon coupling coefficients
$g_{\nu\nu'}$ are small. The use of the zeroth-order approximation 
$\sigma_{\nu\nu'}(\tau)=\sigma_{\nu\nu'}(t)e^{i\omega_{\nu\nu'}(\tau-t)}$
in the expression for $S(\tau)$ [see Eq. (\ref{h11a})] yields 
\begin{equation}
S(\tau)=\sum_{\nu\nu'}g_{\nu\nu'}\sigma_{\nu\nu'}(t)
e^{i\omega_{\nu\nu'}(\tau-t)},
\label{h38}
\end{equation}
which is accurate to first order in the coupling coefficients.
Inserting Eq. (\ref{h38}) into Eq. (\ref{h29}) gives
\begin{equation}
W_{\mathbf{q}}(t)=
\frac{2\pi}{\sqrt{\omega_{\mathbf{q}}}}
\sum_{\nu\nu'}g_{\nu\nu'}\sigma_{\nu\nu'}(t)
\delta_{-}(\omega_{\nu'\nu}-\omega_{\mathbf{q}}),
\label{h39}
\end{equation}
where
\begin{eqnarray}
\delta_{-}(\omega) &=&\lim_{\epsilon\to 0} \frac{1}{2\pi}\int_{-\infty}^{0} 
e^{-i(\omega+i\epsilon)\tau}  \,d\tau
\nonumber\\
&=& \frac{i}{2\pi}\frac{P}{\omega}+\frac{1}{2}\delta(\omega).
\label{h40}
\end{eqnarray}
Here, in order to take into account the effect of adiabatic turn-on of interaction, we have added a small positive parameter $\epsilon$  to the integral and have used the limit 
$t_0\to-\infty$. 
Introducing the notation
\begin{equation} 
K_{\mathbf{q}}=\frac{W_{\mathbf{q}}}{\sqrt{\omega_{\mathbf{q}}}}
=\frac{2\pi}{\omega_{\mathbf{q}}}
\sum_{\nu\nu'}g_{\nu\nu'}\sigma_{\nu\nu'}
\delta_{-}(\omega_{\nu'\nu}-\omega_{\mathbf{q}}),
\label{h41a}
\end{equation} 
we can rewrite
Eq. (\ref{h37}) in the form
\begin{eqnarray}
\lefteqn{\frac{\partial  \langle\mathcal{O}(t)\rangle}{\partial t} 
= \frac{\mathrm{i}}{\hbar}\langle[H_A(t),\mathcal{O}(t)]\rangle}
\nonumber\\&&\mbox{}
+\sum_{\mathbf{q}} (\bar{n}_{\mathbf{q}}+1)
\langle [S(t),\mathcal{O}(t)]K_{\mathbf{q}}(t)+K_{\mathbf{q}}^\dagger(t)[\mathcal{O}(t),S(t)] \rangle
\nonumber\\&&\mbox{}
+\sum_{\mathbf{q}} \bar{n}_{\mathbf{q}}
\langle K_{\mathbf{q}}(t) [\mathcal{O}(t),S(t)] + [S(t),\mathcal{O}(t)]K_{\mathbf{q}}^\dagger(t)\rangle.
\label{h41}
\end{eqnarray}

In order to examine the time evolution of the reduced density 
operator $\rho(t)$ of the atom in the Schr\"odinger picture,
we use the relation $\langle\mathcal{O}(t)\rangle=\mathrm{Tr}[\mathcal{O}(t)\rho(0)]
=\mathrm{Tr}[\mathcal{O}(0)\rho(t)]$, 
transform to arrange the operator $\mathcal{O}(0)$ at the first position in each operator product, and eliminate $\mathcal{O}(0)$. Then, we obtain the Liouville master equation
\begin{eqnarray}
\frac{\partial\rho(t)}{\partial t}&=&
-\frac{\mathrm{i}}{\hbar}[H_A,\rho(t)]
\nonumber\\&&\mbox{}
+\sum_{\mathbf{q}}(\bar{n}_{\mathbf{q}}+1)\{[K_{\mathbf{q}}\rho(t),S]+[S,\rho(t)K_{\mathbf{q}}^\dagger]\}
\nonumber\\&&\mbox{}
+\sum_{\mathbf{q}}\bar{n}_{\mathbf{q}}\{ [S,\rho(t)K_{\mathbf{q}}]+[K_{\mathbf{q}}^\dagger\rho(t),S]\}.
\label{h42}
\end{eqnarray}

Equations (\ref{h41}) and (\ref{h42}) are valid to second order in the coupling coefficients. These  equations allow us to study the time evolution and dynamical characteristics of the atom  interacting with the thermal phonon bath.
We note that Eq. (\ref{h42}) is a particular form of the Zwanzig's generalized
master equation, which can be obtained by the projection operator method \cite{Zwanzig}.

\subsection{Relaxation rates and frequency shifts}
\label{sec:basic B}

We use Eq. (\ref{h42}) to derive an equation for the matrix elements 
$\rho_{jj'}\equiv\langle j|\rho|j'\rangle$
of the reduced density operator of the atom. The result is 
\begin{eqnarray}
\lefteqn{\frac{\partial\rho_{jj'}}{\partial t}=
-i\omega_{jj'}\rho_{jj'}
+\sum_{\nu\nu'}(\gamma_{jj'\nu\nu'}^{e}+\gamma_{jj'\nu\nu'}^{a})\rho_{\nu\nu'}}
\nonumber\\&&\mbox{}
-\sum_{\nu}[(\gamma_{j\nu}^{e}+\gamma_{j\nu}^{a})\rho_{\nu j'}
+(\gamma_{j'\nu}^{e*}+\gamma_{j'\nu}^{a*})\rho_{j\nu}],
\label{h43}
\end{eqnarray}
where the coefficients
\begin{eqnarray}
\gamma_{jj'\nu\nu'}^{e}&=&2\pi\sum_{\mathbf{q}}\frac{\bar{n}_{\mathbf{q}}+1}{\omega_{\mathbf{q}}}g_{j\nu}g_{j'\nu'}
[\delta_{-}(\omega_{\nu j}-\omega_{\mathbf{q}})
\nonumber\\&&\mbox{}
+\delta_{+}(\omega_{\nu' j'}-\omega_{\mathbf{q}})],
\nonumber\\
\gamma_{j\nu}^{e}&=&2\pi\sum_{\mathbf{q}\mu}\frac{\bar{n}_{\mathbf{q}}+1}{\omega_{\mathbf{q}}}g_{j\mu}g_{\nu\mu}
\delta_{-}(\omega_{\nu\mu}-\omega_{\mathbf{q}})\qquad
\label{h44}
\end{eqnarray}
and
\begin{eqnarray}
\gamma_{jj'\nu\nu'}^{a}&=&2\pi\sum_{\mathbf{q}}\frac{\bar{n}_{\mathbf{q}}}{\omega_{\mathbf{q}}}g_{j\nu}g_{j'\nu'}
[\delta_{-}(\omega_{j'\nu'}-\omega_{\mathbf{q}})
\nonumber\\&&\mbox{}
+\delta_{+}(\omega_{j\nu}-\omega_{\mathbf{q}})],
\nonumber\\
\gamma_{j\nu}^{a}&=&2\pi\sum_{\mathbf{q}\mu}\frac{\bar{n}_{\mathbf{q}}}{\omega_{\mathbf{q}}}g_{j\mu}g_{\nu\mu}
\delta_{+}(\omega_{\mu\nu}-\omega_{\mathbf{q}})
\label{h45}
\end{eqnarray}
are the decay parameters associated with the phonon emission
and absorption, respectively.
Here, the notation $\delta_+(\omega)=\delta_-^*(\omega)$ has been used.

Equation (\ref{h43}) describes phonon-induced variations in the populations and coherences
of the translational levels of the atom. 
We analyze the characteristics of the relaxation processes.  
For simplicity of mathematical treatment, we first consider only transitions from discrete levels.
The equation for the diagonal matrix element $\rho_{jj}$ for a discrete level $j$ can be written in the form
\begin{eqnarray}
\frac{\partial\rho_{jj}}{\partial t}&=&
\sum_{\nu}(\gamma_{jj\nu\nu}^{e}+\gamma_{jj\nu\nu}^{a})\rho_{\nu\nu}
\nonumber\\&&\mbox{}
-(\gamma_{jj}^{e}+\gamma_{jj}^{a}+\mathrm{c.c.})\rho_{jj}
\nonumber\\&&\mbox{}
+\mbox{off-diagonal terms}.
\label{h46}
\end{eqnarray}
When the off-diagonal terms are neglected, Eq. (\ref{h46}) reduces to a simple rate equation.
It is clear from Eq. (\ref{h46}) that the rate for the downward transition from an upper level $l$ 
to a lower level $k$ ($k<l$) is 
\begin{equation}
R_{kl}^{e}=\gamma_{kkll}^{e}=
2\pi\sum_{\mathbf{q}}\frac{\bar{n}_{\mathbf{q}}+1}{\omega_{\mathbf{q}}}g_{lk}^2
\delta(\omega_{lk}-\omega_{\mathbf{q}}),
\label{h47}
\end{equation} 
while the rate for the upward transition from a lower level $k$ to an upper level $l$ ($l>k$) is 
\begin{equation}
R_{lk}^{a}=\gamma_{llkk}^{a}=
2\pi\sum_{\mathbf{q}}\frac{\bar{n}_{\mathbf{q}}}{\omega_{\mathbf{q}}}g_{lk}^2
\delta(\omega_{lk}-\omega_{\mathbf{q}}).
\label{h48}
\end{equation}
Equations (\ref{h47}) and (\ref{h48}) are in agreement with
the results of Gortel \textit{et al.} \cite{Gortel}, obtained by using the Fermi golden rule.
We note that $R_{kl}^{e}$ and $R_{lk}^{a}$ with $l\le k$  are mathematically equal to zero
because they have no physical meaning.
For convenience, we introduce the notation $R_{lk}=R_{lk}^e$, $R_{lk}^a$, or 0 for $l<k$, $l>k$, or $l=k$, respectively. It is clear that the off-diagonal coefficients $R_{lk}$ with $l\not=k$ are
the rates of transitions. However, 
the diagonal coefficients $R_{kk}$ have no physical meaning and are mathematically equal to zero.

As seen from Eq. (\ref{h46}), the phonon-mediated depletion rate of a level $k$ is 
$\Gamma_{kk}=2\mathrm{Re}(\gamma_{kk}^{e}+\gamma_{kk}^{a})$.
The explicit expression for this rate is
\begin{eqnarray}
\Gamma_{kk}&=&
2\pi\sum_{\mathbf{q}\mu}\frac{\bar{n}_{\mathbf{q}}+1}{\omega_{\mathbf{q}}}
g_{k\mu}^2\delta(\omega_{k\mu}-\omega_{\mathbf{q}})
\nonumber\\&&\mbox{}
+2\pi\sum_{\mathbf{q}\mu}\frac{\bar{n}_{\mathbf{q}}}{\omega_{\mathbf{q}}}
g_{\mu k}^2\delta(\omega_{\mu k}-\omega_{\mathbf{q}}).
\label{h53}
\end{eqnarray}
We note that $\Gamma_{kk}=\sum_{\mu}(R_{\mu k}^e+R_{\mu k}^a)=\sum_{\mu}R_{\mu k}$.
We can write $\Gamma_{kk}=\Gamma_{kk}^e+\Gamma_{kk}^a$, where
\begin{equation}
\Gamma_{kk}^e=\sum_{\mu<k}R_{\mu k}^e
\label{h53a}
\end{equation}
and
\begin{equation} 
\Gamma_{kk}^a=\sum_{\mu>k}R_{\mu k}^a
\label{h53b}
\end{equation}
are the contributions due to downward transitions (phonon emission) and upward transitions (phonon absorption), respectively.
In the above equations, the summation over $\mu$ can be extended to cover not only the discrete levels but also the continuum levels.

Meanwhile, the equation for the off-diagonal matrix element $\rho_{lk}$ for
a pair of discrete levels $l$ and $k$ can be written in the form
$\partial\rho_{lk}/\partial t=
-(i\omega_{lk}+\gamma_{ll}^{e}+\gamma_{ll}^{a}
+\gamma_{kk}^{e*}+\gamma_{kk}^{a*})\rho_{lk}+\dots$, 
or, equivalently, 
\begin{equation}
\frac{\partial\rho_{lk}}{\partial t}=
-i(\omega_{lk}+\Delta_{lk}-i\Gamma_{lk})\rho_{lk}+\dots  .
\label{h50}
\end{equation}
Here the frequency shift $\Delta_{lk}$ is given by 
\begin{eqnarray}
\lefteqn{\Delta_{lk}=\sum_{\mathbf{q}\mu}\frac{\bar{n}_{\mathbf{q}}+1}{\omega_{\mathbf{q}}}
\bigg(\frac{g_{l\mu}^2}{\omega_{l\mu}-\omega_{\mathbf{q}}}
+\frac{g_{\mu k}^2}{\omega_{\mu k}+\omega_{\mathbf{q}}}\bigg)}
\nonumber\\&&\mbox{}
+\sum_{\mathbf{q}\mu}\frac{\bar{n}_{\mathbf{q}}}{\omega_{\mathbf{q}}}
\bigg(\frac{g_{l\mu}^2}{\omega_{l\mu}+\omega_{\mathbf{q}}}
+\frac{g_{\mu k}^2}{\omega_{\mu k}-\omega_{\mathbf{q}}}\bigg),
\label{h51}
\end{eqnarray}
while the coherence decay rate $\Gamma_{lk}$ is expressed as 
\begin{eqnarray}
\Gamma_{lk}&=&
\pi\sum_{\mathbf{q}\mu}\frac{\bar{n}_{\mathbf{q}}+1}{\omega_{\mathbf{q}}}
\big[g_{l\mu}^2\delta(\omega_{l\mu}-\omega_{\mathbf{q}})
+g_{k\mu}^2\delta(\omega_{k\mu}-\omega_{\mathbf{q}})\big]
\nonumber\\&&\mbox{}
+\pi\sum_{\mathbf{q}\mu}\frac{\bar{n}_{\mathbf{q}}}{\omega_{\mathbf{q}}}
\big[g_{\mu l}^2\delta(\omega_{\mu l}-\omega_{\mathbf{q}})
+g_{\mu k}^2\delta(\omega_{\mu k}-\omega_{\mathbf{q}})\big].
\nonumber\\
\label{h52}
\end{eqnarray}
When we set $l=k$ in Eq. (\ref{h51}), we find $\Delta_{kk}=0$.  
When we set $l=k$ in Eq. (\ref{h52}), we recover Eq. (\ref{h53}).
We note that $\Gamma_{lk}=\sum_{\mu}(R_{\mu l}^e+R_{\mu k}^e+R_{\mu l}^a+R_{\mu k}^a)/2
=\sum_{\mu}(R_{\mu l}+R_{\mu k})/2$.
Comparison between Eqs. (\ref{h52}) and (\ref{h53}) yields the relation 
$\Gamma_{lk}=(\Gamma_{ll}+\Gamma_{kk})/2$. 
We can also write $\Gamma_{lk}=\Gamma_{lk}^e+\Gamma_{lk}^a$, where
$\Gamma_{lk}^e=\sum_{\mu}(R_{\mu l}^e+R_{\mu k}^e)/2$ and 
$\Gamma_{lk}^a=\sum_{\mu}(R_{\mu l}^a+R_{\mu k}^a)/2$ are the contributions
due to downward transitions (phonon emission) and upward transitions (phonon absorption), respectively.
In the above equations, the summation over $\mu$ can be extended to cover not only the discrete levels but also the continuum levels.

We now discuss phonon-mediated transitions from continuum (free) levels.
We start by considering free-to-bound transitions.
For a continuum level $f$ with energy $\mathcal{E}_f>0$, the center-of-mass wave function $\varphi_f(x)$ is normalized per unit energy. In this case, the quantity 
$R_{\nu f}$ becomes the density of the transition rate. 
A free level $f$ can be approximated by a level of a quasicontinuum \cite{Javanainen}. A discretization of the continuum can be realized by using a large box of length $L$ with reflecting boundary
conditions \cite{Luc-Koenig}. We label $E_n$ the energies of the eigenstates in the box
and $\phi_{n}(x)$ the corresponding wave functions. 
Note that such states are standing-wave states \cite{Luc-Koenig,Javanainen}.
The relation between a quasicontinuum-state wave function $\phi_{n_f}(x)$, normalized to unity in the box, and the corresponding continuum-state wave function $\varphi_f(x)$, 
normalized per unit energy, with equal energies $E_{n_f}=\mathcal{E}_f$, is \cite{Luc-Koenig}
\begin{eqnarray}
\varphi_f(x)&\cong&\bigg[\frac{\partial E_{n_f}}{\partial n_f}\bigg]^{-1/2}\phi_{n_f}(x)
\nonumber\\
&\cong&\left(\frac{L}{\pi\hbar}\right)^{1/2}\left(\frac{m}{2E_{n_f}}\right)^{1/4}\phi_{n_f}(x).
\label{h63}
\end{eqnarray}
Consequently, for a single atom initially prepared
in the quasicontinuum standing-wave state $|n_f\rangle=|\phi_{n_f}\rangle$, the rate for the transition 
to an arbitrary bound state $|\nu\rangle$ is approximately given by
\begin{equation}
G_{\nu f}=\frac{\pi\hbar}{L} v_f R_{\nu f},
\label{h64}
\end{equation}
where $v_f=(2\mathcal{E}_f/m)^{1/2}$ is the velocity of the atom in the initial standing-wave state $|f\rangle$.
The phonon-mediated free-to-bound decay rate (adsorption rate)  
is then given by 
\begin{equation}
G_f=\sum_{\nu}G_{\nu f},
\label{h64a}
\end{equation}
where the summation includes only bound levels. 
It is clear from Eq. (\ref{h64}) that, in the continuum limit $L\to\infty$,
the rate $G_{\nu f}$ tends to zero. This is because a free atom can be anywhere in free space and therefore the effect of phonons on a single free atom is negligible. 

In order to get deeper insight into the free-to-bound transition rate density $R_{\nu f}$, we consider a macroscopic atomic ensemble in the thermodynamic limit  \cite{Javanainen}. Suppose that there are $N_0$ atoms in a volume with a large length $L$ and a transverse cross section area $S_0$. Assume that all 
the atoms are in the same quasicontinuum  state $|n_f\rangle$ and 
interact with the dielectric independently.
The rate for the transitions of the atoms from 
the quasicontinuum state $|n_f\rangle$ to an arbitrary bound state $|\nu\rangle$, 
defined as the time derivative of the number of atoms in the state $|\nu\rangle$,
is $D_{\nu f}=N_0G_{\nu f}$.
In order to get the rate for the continuum state $|f\rangle$, we need to take  
the thermodynamical limit, where 
$L\to\infty$ and $N_0\to\infty$ but $N_0/L$ remains constant.
Then, the rate for the transitions of the atoms from 
the continuum state $|f\rangle$ to an arbitrary bound state $|\nu\rangle$ is given by
$D_{\nu f}=\pi\hbar \rho_0S_0v_f R_{\nu f}=2\pi\hbar\mathcal{N}_f R_{\nu f}$.
Here, $\rho_0=N_0/LS_0$ is the atomic number density 
and $\mathcal{N}_f=\rho_0S_0v_f/2$  
is the number of atoms incident into the dielectric surface per unit time.
It is clear that the transition rate $D_{\nu f}$ is proportional to 
the incidence rate $\mathcal{N}_f$ as well as the transition rate density $R_{\nu f}$.
We emphasize that $D_{\nu f}$ is a characteristics for a macroscopic atomic ensemble 
in the thermodynamic limit while $G_{\nu f}$ is a measure for a single atom.
When the length of the box, $L$, and the number of atoms, $N_0$, are finite, the dynamics of the atoms
cannot be described by the free-to-bound rate $D_{\nu f}$ directly. Instead, we must
use the transition rate per atom $G_{\nu f}=D_{\nu f}/N_0$, which depends on the length $L$ of the box
that contains the free atoms [see Eq. (\ref{h64})].

In a thermal gas, the atoms have different velocities and, therefore, different energies. 
For a thermal Maxwell-Boltzmann gas with temperature $T_0$, the distribution of the kinetic energy $\mathcal{E}_f$ of the atomic center-of-mass motion
along the $x$ direction is
\begin{equation}
P(\mathcal{E}_f)=\frac{1}{\sqrt{\pi k_BT_0}}
\frac{e^{-\mathcal{E}_f/k_BT_0}}{\sqrt{\mathcal{E}_f}}.
\label{h66}
\end{equation}
The transition rate to an arbitrary bound state $|\nu\rangle$ is then given by 
$G_{\nu T_0}=\int_0^{\infty}G_{\nu f}P(\mathcal{E}_f)\, d\mathcal{E}_f$, i.e.
\begin{equation}
G_{\nu T_0}
=\frac{\lambda_D}{L}
\int_0^{\infty}e^{-\mathcal{E}_f/k_BT_0}R_{\nu f}d\mathcal{E}_f,
\label{h67}
\end{equation}
where $\lambda_D=(2\pi\hbar^2/mk_BT_0)^{1/2}$ is the thermal de Broglie wavelength.
The phonon-mediated free-to-bound decay rate (adsorption rate) is given by 
\begin{equation}
G_{T_0}=\sum_{\nu}G_{\nu T_0}=\int_0^{\infty}G_f P(\mathcal{E}_f)\, d\mathcal{E}_f.
\label{h67a}
\end{equation}
In the above equation, the summation over $\nu$ includes only bound levels. 
Note that Eq. (\ref{h67}) is in qualitative agreement with the results of 
Refs. \cite{Oria2006,Javanainen}. 

It is easy to extend the above results to the case of free-to-free transitions.
Indeed, it can be shown that the density of the rate for the transition from a quasicontinuum state $|n_f\rangle$, which corresponds to a free state $|f\rangle$, to a different free state $|f'\rangle$ is given by  
\begin{equation}
Q_{f'f}=\frac{\pi\hbar}{L} v_f R_{f'f}.
\label{h68}
\end{equation}
For convenience, we introduce the notation 
$Q_{f'f}^e=Q_{f'f}$ or 0 for $\mathcal{E}_{f'}<\mathcal{E}_f$ or 
$\mathcal{E}_{f'}\geq \mathcal{E}_f$, respectively, and 
$Q_{f'f}^a=Q_{f'f}$ or 0 for $\mathcal{E}_{f'}>\mathcal{E}_f$ or 
$\mathcal{E}_{f'}\leq \mathcal{E}_f$, respectively.
Then, we have $Q_{f'f}=Q_{f'f}^e$, 0, or $Q_{f'f}^a$
for $\mathcal{E}_{f'}<\mathcal{E}_f$, $\mathcal{E}_{f'}=\mathcal{E}_f$, 
or $\mathcal{E}_{f'}>\mathcal{E}_f$, respectively.
The downward (phonon-emission) and upward (phonon-absorption) free-to-free 
decay rates for the free state $|f\rangle$ are given by
\begin{equation}
Q_f^e=\int_0^{\mathcal{E}_f}Q_{f'f}^e d\mathcal{E}_{f'}
\label{h68a}
\end{equation}
and 
\begin{equation}
Q_f^a=\int_{\mathcal{E}_f}^{\infty} Q_{f'f}^a d\mathcal{E}_{f'},
\label{h68b}
\end{equation}
respectively. The total free-to-free decay rate for the free state $|f\rangle$ is
$Q_f=Q_f^e+Q_f^a=\int_0^{\infty} Q_{f'f} d\mathcal{E}_{f'}$.

For a thermal gas, we need to replace the transition rate density 
$Q_{f'f}$ and the decay rate $Q_f$ by 
$Q_{f'T_0}=\int_0^{\infty}Q_{f'f} P(\mathcal{E}_f)\, d\mathcal{E}_f$ and 
$Q_{T_0}=\int_0^{\infty}Q_f P(\mathcal{E}_f)\, d\mathcal{E}_f$, respectively, which are the
averages of  $Q_{f'f}$ and $Q_f$, respectively, with respect to the energy distribution $P(\mathcal{E}_f)$ of the initial state. Like in the other cases,  
we have $Q_{f'T_0}=Q_{f'T_0}^e+Q_{f'T_0}^a$
and $Q_{T_0}=Q_{T_0}^e+Q_{T_0}^a$, where
\begin{eqnarray}
Q_{f'T_0}^e&=&\int_{\mathcal{E}_{f'}}^{\infty}Q_{f'f}^e P(\mathcal{E}_f)\, d\mathcal{E}_f,\nonumber\\
Q_{f'T_0}^a&=&\int_0^{\mathcal{E}_{f'}}Q_{f'f}^a P(\mathcal{E}_f)\, d\mathcal{E}_f
\label{h68c}
\end{eqnarray} 
are the downward and upward transition rate densities and 
\begin{eqnarray}
Q_{T_0}^e&=&\int_0^{\infty}Q_f^e P(\mathcal{E}_f)\, d\mathcal{E}_f,
\nonumber\\
Q_{T_0}^a&=&\int_0^{\infty}Q_f^a P(\mathcal{E}_f)\, d\mathcal{E}_f
\label{h68d}
\end{eqnarray}
are the downward and upward decay rates. The thermal decay rates $Q_{T_0}^e$ and $Q_{T_0}^a$
describe the cooling and heating processes, respectively.  
It can be easily shown that $Q_{T_0}^e<Q_{T_0}^a$, $Q_{T_0}^e>Q_{T_0}^a$,
and $Q_{T_0}^e=Q_{T_0}^a$ when $T_0<T$, $T_0>T$, and $T_0=T$, respectively. 
The relation $Q_{T_0}^e<Q_{T_0}^a$ ($Q_{T_0}^e>Q_{T_0}^a$), obtained
for $T_0<T$ ($T_0>T$), indicates the dominance of heating (cooling) of free atoms by the surface.

\subsection{Relaxation rates and frequency shifts in the framework of the Debye model}
\label{sec:basic C}

In order to get insight into the relaxation rates and frequency shifts, we approximate them using the Debye model for phonons. In this  model, the phonon frequency $\omega_{\mathbf{q}}$ is related to the phonon wave number $q$ as 
$\omega_{\mathbf{q}}=vq$, where $v$ is the sound velocity. Furthermore, the summation over the first Brillouin zone is replaced by an integral over a sphere of radius 
$q_D=(6\pi^2N/V)^{1/3}$, where $V$ is the volume of the solid. The Debye frequency and the Debye temperature are given by
$\omega_D=vq_D$ and $T_D=\hbar\omega_D/k_B$, respectively. 
For fused silica, we have $v=5.96$ km/s, $NM/V=2.2$ g/cm$^3$,
and $M=9.98\times 10^{-26}$ kg \cite{Agrawal}. 
Using these parameters, we find $q_D=109.29\times10^6$ cm$^{-1}$, 
$\omega_D=10.4$ THz, and $T_D=498$ K.
In order to perform the summation over phonon states in the framework of the Debye model, 
we invoke the thermodynamic limit, i.e., replace
\begin{equation}
\sum_{\mathbf{q}}\dots
=\frac{V}{8\pi^3}\int\limits_{|\mathbf{q}|\leq q_D} \dots d\mathbf{q}
=\frac{3N}{\omega_D^3}\int\limits_0^{\omega_D}\dots\omega_{\mathbf{q}}^2 d\omega_{\mathbf{q}}.
\label{h54}
\end{equation} 
Then, for transitions between an upper level $l$ and a lower level $k$, 
where $0<\omega_{lk}<\omega_D$, 
Eqs. (\ref{h47}) and (\ref{h48}) yield
\begin{equation}
R_{kl}^{e}=\frac{3\pi}{M\hbar\omega_D^3}(\bar{n}_{lk}+1)\omega_{lk} F_{lk}^2
\label{h57}
\end{equation}
and 
\begin{equation}
R_{lk}^{a}=\frac{3\pi}{M\hbar\omega_D^3}\bar{n}_{lk}\omega_{lk} F_{lk}^2.
\label{h58}
\end{equation}
Here, $\bar{n}_{lk}$ is given by Eq. (\ref{h22}) with $\omega_{\mathbf{q}}$ 
replaced by $\omega_{lk}$.
We emphasize that, according to Eqs. (\ref{h57}) and  (\ref{h58}), 
the phonon-emission rate $R_{kl}^{e}$ and the phonon-absorption rate $R_{lk}^{a}$
depend not only on the matrix element $F_{lk}$ of the force but also
on the translational transition frequency $\omega_{lk}$. 
The frequency dependences of the transition rates are comprised of
the frequency dependences of the mean phonon number $\bar{n}_{lk}$, 
the phonon mode density $3N\omega_{lk}^2/\omega_D^3$,
and the matrix element $F_{lk}=-U_{lk}'=-m\omega_{lk}^2x_{lk}$ of the force. An additional factor comes from
the presence of the phonon frequency in Eq. (\ref{h5}) for the surface displacement and, 
consequently, in the atom--phonon interaction Hamiltonian (\ref{h6}). 
It is clear that an increase in the phonon frequency leads to a decrease in the mean phonon
number and an increase in the phonon mode density. The matrix element of the force
usually first increases and then decreases with increasing phonon frequency.
Due to the existence of several competing factors, the frequency dependences of the transition rates are rather complicated.
They usually first increase and then decrease with increasing phonon frequency.
We note that, for transitions with $\omega_{lk}>\omega_D$, 
we have $R_{kl}^{e}=R_{lk}^{a}=0$. 

We conclude this section by noting that
the use of Eq. (\ref{h54}) in Eq. (\ref{h51}) yields the frequency shift
\begin{equation}
\Delta_{lk}=\Delta_{lk}^{(0)}+\Delta_{lk}^{(T)},
\end{equation} 
where
\begin{equation}
\Delta_{lk}^{(0)}=
\frac{3}{2M\hbar\omega_D^3}\sum_{\mu}
\int\limits_0^{\omega_D}
\bigg(\frac{F_{l\mu}^2}{\omega_{l\mu}-\omega}
+\frac{F_{\mu k}^2}{\omega_{\mu k}+\omega}\bigg)\omega d\omega 
\label{h61}
\end{equation}
and
\begin{equation}
\Delta_{lk}^{(T)}=
\frac{3}{M\hbar\omega_D^3}\sum_{\mu}
\int\limits_0^{\omega_D}
\bigg(
\frac{\omega_{l\mu} F_{l\mu}^2}{\omega_{l\mu}^2-\omega^2}
+\frac{\omega_{\mu k} F_{\mu k}^2}{\omega_{\mu k}^2-\omega^2}\bigg)\bar{n}_{\omega}\omega d\omega 
\label{h62}
\end{equation}
are the zero- and finite-temperature contributions, respectively.
In Eq. (\ref{h62}), $\bar{n}_{\omega}$
is given by Eq. (\ref{h22}) with $\omega_{\mathbf{q}}$ replaced by $\omega$.

\section{Numerical results and discussions}
\label{sec:numerical}

In this section, we present the numerical results based on the analytical expressions derived in the previous section for the phonon-mediated relaxation rates of the translational levels of the atom. In particular, we use Eqs. (\ref{h57}) 
and (\ref{h58}), obtained in the framework of the Debye model, for our numerical calculations. We consider transitions from bound states as well as free states. The transitions from bound states to other translational levels occur in the case where the atom is initially already adsorbed or trapped near the surface. The transitions from free states to other translational levels occur in the processes of adsorbing, heating, and cooling of free atoms by the surface. Due to the difference in physics of the initial situations, we study the transitions from bound and free states separately.

\subsection{Transitions from bound states}

\begin{figure}[tbh]
\begin{center}
 \end{center}
\caption{Phonon-emission rates $R_{\nu'\nu}^e$ from the vibrational levels (a) $\nu=280$ and (b) $\nu=120$
to other levels $\nu'$ as functions of the lower-level energy $\mathcal{E}_{\nu'}$. 
The arrows mark the initial states.
The parameters of the solid are 
$M=9.98\times 10^{-26}$ kg and $\omega_D=10.4$ THz.
The temperature of the phonon bath is $T=300$ K. 
Other parameters are as in Fig.~\ref{fig1}.
}
\label{fig3}
\end{figure}

\begin{figure}[tbh]
\begin{center}
 \end{center}
\caption{Phonon-absorption rates $R_{\nu'\nu}^a$ from the vibrational levels (a) $\nu=280$ and (b) $\nu=120$
to other levels $\nu'$ as functions of the upper-level energy $\mathcal{E}_{\nu'}$. The left (right) panel in each row corresponds to bound-to-bound (bound-to-free) transitions.
The arrows mark the initial states.
The parameters used are as in Fig.~\ref{fig3}. 
The temperature of the phonon bath is $T=300$ K. 
}
\label{fig4}
\end{figure}

We start from a given bound level and calculate the rates of phonon-mediated atomic transitions, both downward and upward. The  profiles of the phonon-emission (downward-transition) rate $R_{\nu'\nu}^e$ [see Eq. (\ref{h57})] and the phonon-absorption (upward-transition) rate $R_{\nu'\nu}^a$ [see Eq. (\ref{h58})] are shown in Figs.~\ref{fig3} and \ref{fig4}, respectively. 
The upper (lower) part  of each of these figures corresponds to the case of the initial level $\nu=280$ ($\nu=120$), with energy $\mathcal{E}_{\nu}=-156$ MHz ($\mathcal{E}_{\nu}=-8.4$ THz). The left (right) panel of Fig.~\ref{fig4} corresponds to bound-to-bound (bound-to-free) upward transitions. The temperature of the surface is assumed to be $T=300$ K.
As seen from Figs.~\ref{fig3} and \ref{fig4}, 
the transition rates have pronounced localized profiles.
Due to the competing effects of the mean phonon number, 
the phonon mode density, and the matrix element of the force, 
the transition rates usually first increase and then decrease with increasing phonon frequency.
It is clear from a comparison
of Figs.~\ref{fig3}(a) and \ref{fig3}(b) and also a comparison of Figs.~\ref{fig4}(a) and \ref{fig4}(b) that transitions from shallow levels have probabilities orders
of magnitude lower than those from deeper levels. The main reason is that
the wave functions of the shallow states are spread further
away from the surface than those for the deep states.
Due to this difference, the effects of the surface vibrations are weaker for the shallow levels than for the deep levels. Another pertinent feature that should be noted from the figure is the following: Since transition frequencies involved are large, they may overshoot the Debye frequency $\omega_D=10.4$ THz, leading to a cutoff on the lower (higher) side of the frequency axis for the emission (absorption) curve.

In order to see the overall effect of the individual transition rates shown above,
we add them up. First we examine the phonon-absorption rates of bound levels. The total phonon-absorption rate $\Gamma_{\nu\nu}^a$ of a bound level $\nu$ is the sum of the individual absorption rates $R_{\mu\nu}^a$ over all the upper levels $\mu$, both bound and free [see Eq. (\ref{h53b})].  
We plot in Fig.~\ref{fig5} the contributions to $\Gamma_{\nu\nu}^a$ from two types of transitions, bound-to-bound and bound-to-free (desorption) transitions. 
The solid curve of the figure shows that the bound-to-bound phonon-absorption rate is large (above
$10^{10}$ s$^{-1}$) for deep and intermediate levels. However, it reduces dramatically with increasing $\nu$ in the region of large $\nu$ and becomes very small 
(below $10^{-5}$ s$^{-1}$) for shallow levels. 
Meanwhile, the dashed curve of Fig.~\ref{fig5} shows that the bound-to-free phonon-absorption rate (i.e., the desorption rate) is zero for deep levels, since the energy required for the transition is greater than the Debye energy \cite{Oria2006}.  
However, the desorption rate is substantial (above $10^5$ s$^{-1}$) for intermediate and shallow levels.
Thus, the total phonon-absorption rate $\Gamma_{\nu\nu}^a$ is mainly determined by
the bound-to-bound transitions in the case of deep levels and
by the bound-to-free transitions in the case of shallow levels.
One of the reasons for the dramatic reduction of the bound-to-bound phonon-absorption rate in the region of shallow levels is that the number of upper bound levels $\mu$ becomes small.
The second reason is that the frequency of each individual transition becomes small,
leading to a decrease of the phonon mode density. The third reason is that
the center-of-mass wave functions of shallow levels are spread far away from the surface,
leading to a reduction of the effect of phonons on the atom.

Unlike the bound-to-bound phonon-absorption rate, the bound-to-free phonon-absorption rate is substantial in the region of shallow levels. This is because the free-state spectrum is continuous and the range of the bound-to-free transition frequency can be large (up to the Debye frequency $\omega_D=10.4$ THz). The gradual reduction of the bound-to-free phonon-absorption rate in the region of shallow levels is mainly due to the reduction of the time that the atom spends in the proximity of the surface.

\begin{figure}[tbh]
\begin{center}
 \end{center}
\caption{Contributions of bound-to-bound (solid curve)  and bound-to-free (dashed curve) transitions to the total phonon-absorption rate $\Gamma_{\nu\nu}^a$  versus the vibrational quantum number $\nu$ of the initial level. 
The parameters used are as in Fig.~\ref{fig3}. 
The temperature of the phonon bath is $T=300$ K. 
}
\label{fig5}
\end{figure}

The total phonon-emission rate $\Gamma_{\nu\nu}^e$ [see Eq. (\ref{h53a})] and the total phonon-absorption rate 
$\Gamma_{\nu\nu}^a$ [see Eq. (\ref{h53b})] are shown in Fig.~\ref{fig6} by the solid and dashed curves, respectively. It is clear from the figure that emission is comparable to but slightly stronger than absorption. Such a dominance is due to the fact that phonon emission moves the atom to a center-of-mass state closer to the surface while phonon absorption changes the atomic state in the opposite direction (see Figs.~\ref{fig1} and \ref{fig2}). Our results for the rates are in good qualitative agreement with the results of Oria \textit{et al.}, albeit with the Morse potential \cite{Oria2006}. We stress that we include a large number of vibrational levels as a consequence of the deep silica--cesium potential. Note that the earlier work on this theme involved much fewer levels \cite{Oria2006}. 

\begin{figure}[tbh]
\begin{center}
 \end{center}
\caption{
Phonon-emission decay rate $\Gamma_{\nu\nu}^e$ (solid lines) 
and phonon-absorption decay rate $\Gamma_{\nu\nu}^a$ (dashed lines) 
of a bound level as functions of the vibrational quantum number $\nu$. 
The inset shows the rates in the linear scale to highlight
the differences in the dissociation limit. The parameters used are as in Fig.~\ref{fig3}. 
The temperature of the phonon bath is $T=300$ K. 
}
\label{fig6}
\end{figure}

\begin{figure}[tbh]
\begin{center}
 \end{center}
\caption{Same as in Fig.~\ref{fig6} except that $T=30$ K. 
}
\label{fig7}
\end{figure}

We next study the effect of temperature on the decay rates. The results for
the phonon-mediated decay rates for $T=30$ K are shown in Fig.~\ref{fig7}.
In contrast to Fig.~\ref{fig6}, the absorption rate is now much smaller than
the corresponding emission rate for both shallow and deep levels.
Thus, while it is difficult to distinguish the two log-scale curves for deep and shallow levels at room temperature (see Fig.~\ref{fig6}), they are well resolved
at low temperature.

\subsection{Transitions from free states}

We now calculate the rates for transitions from free states to other levels.
We first examine free-to-bound transitions, which correspond to the adsorption process. 
According to Eq. (\ref{h64}), the free-to-bound (more exactly, quasicontinuum-to-bound) transition rate $G_{\nu f}$ depends not only on the continuum-to-bound transition rate density $R_{\nu f}$ but also on the length $L$ of the free-atom quantization box. To be specific, we use in our numerical calculations the value $L=1$ mm, which is a typical size of atomic clouds in magneto-optical traps \cite{coolingbook}. 

\begin{figure}[tbh]
\begin{center}
 \end{center}
\caption{Free-to-bound transition rates $G_{\nu f}$ for transitions 
from the free plane-wave states with energies (a) 
$\mathcal{E}_f=2$ MHz and (b) $\mathcal{E}_f=3.1$ THz
to bound levels $\nu$ as functions of the bound-level energy $\mathcal{E}_{\nu}$. 
The arrows mark the energies of the initial free states.
The insets show $G_{\nu f}$ on the log scale versus $\mathcal{E}_{\nu}$ 
in the range from $-200$ MHz to $-0.2$ MHz 
to highlight the rates to shallow bound levels.
The length of the free-atom quantization box is $L=1$ mm.
The temperature of the phonon bath is $T=300$ K. 
Other parameters are as in Fig.~\ref{fig3}. 
}
\label{fig8}
\end{figure}

We plot in Fig.~\ref{fig8} the free-to-bound transition rate $G_{\nu f}$ 
[see Eq. (\ref{h64})] as a function
of the vibrational quantum number $\nu$. 
The upper (lower) part  of the figure corresponds to the case of 
the initial-state energy $\mathcal{E}_f=2$ MHz ($\mathcal{E}_f=3.1$ THz),
which is close to the average kinetic energy per atom in an ideal gas 
with temperature $T_0=200$ $\mu$K ($T_0=300$ K). We observe that
the free-to-bound transition rate first increases and then decreases with increasing transition frequency $\omega_{f\nu}=(\mathcal{E}_f-\mathcal{E}_{\nu})/\hbar$. 
Such behavior results from the competing effects of the mean phonon number, 
the phonon mode density, and the matrix element of the force,
like in the case of bound-to-bound transitions (see Fig.~\ref{fig3}).
We also see a cutoff of the transition frequency, which is associated with the Debye frequency. Comparison of Figs.~\ref{fig8}(a) and \ref{fig8}(b) shows that the transitions from low-energy free states have probabilities orders of magnitude smaller than those from high-energy free states. 
One of the reasons is that the transition rate $G_{\nu f}$ is proportional to the velocity 
$v_f=(2\mathcal{E}_f/m)^{1/2}$ [see Eq. (\ref{h64})].     
The dependence of the transition rate density 
$R_{\nu f}$ on the transition frequency $\omega_{f\nu}$ also plays an important role. 
Because of this, the rates for the transitions from  
low-energy free states  to shallow bound levels are very small
[see the inset of Fig.~\ref{fig8}(a)]. 
 
\begin{figure}[tbh]
\begin{center}
 \end{center}
\caption{Free-to-bound decay rate $G_f$ as a function of the free-state energy $\mathcal{E}_f$. 
The inset highlights the magnitude and profile of the decay rate for $\mathcal{E}_f$ 
in the range  from 0 to 20 MHz. 
The temperature of the phonon bath is $T=300$ K. 
Other parameters are as in Fig.~\ref{fig8}. 
}
\label{fig9}
\end{figure}

We show in Fig.~\ref{fig9} the free-to-bound decay rate $G_f$ [see Eq. (\ref{h64a})], which is a characteristic of the adsorption process,
as a function of the free-state energy $\mathcal{E}_f$. 
We see that $G_f$ first  increases and then 
decreases with increasing  $\mathcal{E}_f$.
The increase of $G_f$ with increasing  $\mathcal{E}_f$ in the region of small $\mathcal{E}_f$ (see the inset) is mainly due to the increase in 
the atomic incidence velocity $v_f$. In this region, we have 
$G_f\propto v_f\propto \sqrt{\mathcal{E}_f}$ [see Eqs. (\ref{h64}) and (\ref{h64a})].
For $\mathcal{E}_f$ in the range from 0 to 20 MHz, which is typical for atoms in magneto-optical traps, the maximum value of $G_{f}$ is on the order of $10^4$ s$^{-1}$
(see the inset of Fig.~\ref{fig9}).
Such free-to-bound (adsorption) rates are several orders of magnitude smaller than the bound-to-free (desorption) rates (see the dashed curve in Fig.~\ref{fig5}).
The decrease of $G_f$ with increasing  $\mathcal{E}_f$ in the region of large $\mathcal{E}_f$ is mainly due to the reduction of the atom--phonon coupling coefficients.

\begin{figure}[tbh]
\begin{center}
 \end{center}
\caption{Free-to-bound transition rates $G_{\nu T_0}$ for transitions 
from the thermal states with temperatures (a) 
$T_0=200$ $\mu$K and (b) $T_0=300$ K
to bound levels $\nu$ as functions of the bound-level energy $\mathcal{E}_{\nu}$. 
The insets show $G_{\nu T_0}$ on the log scale versus $\mathcal{E}_{\nu}$ 
in the range from $-200$ MHz to $-0.2$ MHz 
to highlight the rates to shallow bound levels.
The temperature of the phonon bath is $T=300$ K. 
Other parameters are as in Fig.~\ref{fig8}. 
}
\label{fig10}
\end{figure}

\begin{figure}[tbh]
\begin{center}
 \end{center}
\caption{Free-to-bound decay rate $G_{T_0}$ 
as a function of the atomic temperature $T_0$ in the ranges 
(a) from 100 $\mu$K to 400 $\mu$K and (b) from  50 K to 350 K. 
The temperature of the phonon bath is $T=300$ K. 
Other parameters are as in Fig.~\ref{fig8}. 
}
\label{fig11}
\end{figure}

In a thermal gas, the adsorption process is characterized by the transition rate $G_{\nu T_0}$ [see Eq. (\ref{h67})] 
and the decay rate $G_{T_0}$ [see Eq. (\ref{h67a})], which are the averages of the free-to-bound transition rate 
$G_{\nu f}$ and the free-to-bound decay rate $G_f$, respectively, 
over the free-state energy distribution (\ref{h66}). 
We plot the free-to-bound transition rate $G_{\nu T_0}$ and the 
free-to-bound decay rate $G_{T_0}$
in Figs.~\ref{fig10} and \ref{fig11}, respectively. 
Comparison between Figs.~\ref{fig10}(a) and \ref{fig9}(a) shows that
the transition rates from low-temperature thermal states and low-energy free states look  quite similar to each other. The reason is that the spread of the energy distribution
is not substantial in the case of low temperatures.
The spread of the energy distribution
is however substantial in the case of high temperatures, leading to the softening of the cutoff frequency effect [compare Fig.~\ref{fig10}(b) with Fig.~\ref{fig9}(b)].
Figure \ref{fig11} shows that the free-to-bound decay rate $G_{T_0}$
first increases and then reduces with increasing atomic temperature $T_0$.
For $T_0$ in the range from 100 $\mu$K to 400 $\mu$K, which is typical for atoms in magneto-optical traps, the maximum value of $G_{T_0}$ is on the order of $10^4$ s$^{-1}$
[see Fig.~\ref{fig11}(a)].
Such free-to-bound (adsorption) rates are several orders of magnitude smaller than the bound-to-free (desorption) rates (see the dashed curve in Fig.~\ref{fig5}).
Figure \ref{fig11}(a) shows that, in the region of low atomic temperature $T_0$, 
one has $G_{T_0}\propto \sqrt{T_0}$, 
in agreement with the asymptotic behavior of Eqs. (\ref{h67}) and (\ref{h67a}).

\begin{figure}[tbh]
\begin{center}
 \end{center}
\caption{Free-to-free transition rate densities $Q_{f'f}$ for the upward (solid lines) and
downward (dashed lines) transitions 
from the free states $|f\rangle$ with energies (a) 
$\mathcal{E}_f=2$ MHz and (b) $\mathcal{E}_f=3.1$ THz
to other free states $|f'\rangle$ as functions of the final-level energy $\mathcal{E}_{f'}$. 
The arrows mark the energies of the initial free states.
The inset in part (a) shows $Q_{f'f}$ versus 
$\mathcal{E}_{f'}$ in the range from 0 to 4 MHz to highlight the small magnitude of the rate density for downward transitions (dashed line).
The temperature of the phonon bath is $T=300$ K. 
Other parameters are as in Fig.~\ref{fig8}. 
}
\label{fig12}
\end{figure}

We now examine free-to-free transitions, both upward and downward, which corresponding to the heating and cooling processes of free atoms by the surface. We plot in Fig.~\ref{fig12} the free-to-free transition rate density $Q_{f'f}$ [see Eq. (\ref{h68})] as a function of the final-level energy 
$\mathcal{E}_{f'}$. The upper (lower) part  of the figure corresponds to the case of 
the initial-state energy $\mathcal{E}_f=2$ MHz ($\mathcal{E}_f=3.1$ THz),
which is close to the average kinetic energy per atom in an ideal gas 
with temperature $T_0=200$ $\mu$K ($T_0=300$ K). 
The rate densities are shown for the upward (phonon-absorption)  and downward (phonon-emission) transitions by the solid and dashed lines, respectively.
The figure shows that the free-to-free transition rate density increases or decreases with increasing transition frequency if the latter is not too large or is large enough, respectively. We also observe a signature of the Debye cutoff of the phonon frequency. Comparison of Figs.~\ref{fig12}(a) and \ref{fig12}(b) shows that transitions from low-energy free states have probabilities orders of magnitude smaller than those from high-energy free states. Figure \ref{fig12}(a) and its inset show that, 
when the energy of the free state is low, the free-to-free downward (cooling) transition rate is very small as compared to the free-to-free upward (heating) transition rate.

\begin{figure}[tbh]
\begin{center}
 \end{center}
\caption{Free-to-free upward and downward decay rates $Q_f^a$ (solid lines) and $Q_f^e$ (dashed lines) as functions of the energy $\mathcal{E}_f$ of the initial free state. The insets highlight the magnitudes and profiles of the decay rates for $\mathcal{E}_f$
in the range  from 0 to 20 MHz. 
The temperature of the phonon bath is $T=300$ K. 
Other parameters are as in Fig.~\ref{fig8}. 
}
\label{fig13}
\end{figure}

We show in Fig.~\ref{fig13} the free-to-free upward (phonon-absorption) and downward (phonon-emission) decay rates $Q_f^a$ [see Eq. (\ref{h68b})] 
and $Q_f^e$ [see Eq. (\ref{h68a})]
as functions of the free-state energy $\mathcal{E}_f$. 
We observe that $Q_f^a$ and $Q_f^e$ increase with increasing  $\mathcal{E}_f$ in the range
from 0 to 8 THz.
The increase of $Q_f^a$ with increasing  $\mathcal{E}_f$ in the region of small $\mathcal{E}_f$ (see the left inset) is mainly due to the increase in 
the atomic incidence velocity $v_f$. In this region, we have 
$Q_f^a\propto v_f\propto \sqrt{\mathcal{E}_f}$ [see Eqs. (\ref{h68}) and (\ref{h68b})].
The increase of $Q_f^e$ with increasing  $\mathcal{E}_f$ in the region of small $\mathcal{E}_f$ (see the right inset) is due to not only the increase in 
the atomic incidence velocity $v_f$ [see Eq. (\ref{h68})] but also the increase of the transition rate density $Q_{f'f}^e$ and the increase of 
the integration interval $(0,\mathcal{E}_f)$ [see Eq. (\ref{h68a})]. 
In this region, the dependence of $Q_f^e$ on the energy $\mathcal{E}_f$ 
is of higher order than $\mathcal{E}_f^{3/2}$.
The left inset of Fig.~\ref{fig13} shows that, for $\mathcal{E}_f$ in the range from 0 to 20 MHz, the maximum value of $Q_f^a$ is on the order of $10^4$ s$^{-1}$.
Such free-to-free upward (heating) decay rates are comparable to but about two times smaller than the corresponding free-to-bound (adsorption) decay rates (see the inset of Fig.~\ref{fig9}).
Meanwhile, the right inset of Fig.~\ref{fig13} shows that, 
in the region of small $\mathcal{E}_f$, the free-to-free downward (cooling) decay rate $Q_f^e$ is very small.

\begin{figure}[tbh]
\begin{center}
 \end{center}
\caption{Free-to-free transition rate densities $Q_{f T_0}^a$ for upward transitions (solid lines) and $Q_{f T_0}^e$ for downward transitions (dashed lines) 
from the thermal states with temperatures (a) 
$T_0=200$ $\mu$K and (b) $T_0=300$ K
to free levels $f$ as functions of the free-level energy $\mathcal{E}_{f}$. 
The inset in part (a) shows the rate densities versus $\mathcal{E}_{f}$ 
in the range from 0 to 8 MHz 
to highlight the small magnitude of $Q_{f T_0}^e$ (dashed line).
The temperature of the phonon bath is $T=300$ K. 
Other parameters are as in Fig.~\ref{fig8}. 
}
\label{fig14}
\end{figure}

\begin{figure}[tbh]
\begin{center}
 \end{center}
\caption{Free-to-free decay rates $Q_{T_0}^a$ (solid lines) and $Q_{T_0}^e$(dashed lines)  
for upward and downward transitions, respectively, 
as functions of the atomic temperature $T_0$ in the ranges 
(a) from 100 $\mu$K to 400 $\mu$K and (b) from  50 K to 350 K. 
For comparison, the free-to-bound decay rate $G_{T_0}$ is re-plotted
from Fig.~\ref{fig11} by the dotted lines.
The temperature of the phonon bath is $T=300$ K. 
Other parameters are as in Fig.~\ref{fig8}. 
}
\label{fig15}
\end{figure}

In the case of a thermal gas, the phonon-mediated heat transfer 
between the gas and the surface is characterized by the free-to-free transition rate densities $Q_{f T_0}^a$ and $Q_{f T_0}^e$ [see Eqs. (\ref{h68c})] and the free-to-free decay rates $Q_{T_0}^a$and  $Q_{T_0}^e$ [see Eqs. (\ref{h68d})].
We plot the free-to-free transition rate densities $Q_{f T_0}^a$ and $Q_{f T_0}^e$ in Fig.~\ref{fig14}. Comparison between Figs.~\ref{fig14}(a) and \ref{fig12}(a) shows that
the transition rate densities from low-temperature thermal states and low-energy free states are quite similar to each other. The spread of the initial-state energy distribution
is not substantial in this case. However, the energy spread of the initial state
is substantial in the case of high temperatures, concealing the cutoff frequency effect [compare Fig.~\ref{fig14}(b) with Fig.~\ref{fig12}(b)]. We display
the free-to-free decay rates $Q_{T_0}^a$ and $Q_{T_0}^e$ in Fig.~\ref{fig15}. 
The solid and dashed lines correspond to the upward (heating) and downward (cooling) transitions, 
respectively. For comparison, the free-to-bound decay rate 
(adsorption rate) $G_{T_0}$ is re-plotted
from Fig.~\ref{fig11} by the dotted lines.
We observe that, for $T_0$ in the range from 100 $\mu$K to 400 $\mu$K 
[see Fig. \ref{fig15}(a)], the adsorption rate $G_{T_0}$ (dotted line) is about two times larger than the heating rate $Q_{T_0}^a$ (solid line), while the cooling rate $Q_{T_0}^e$ (dashed line) is negligible. Figure \ref{fig15}(a) shows that, in the region of low atomic temperatures, 
one has $Q_{T_0}\cong Q_{T_0}^a\propto \sqrt{T_0}$, in agreement with the asymptotic
behavior of expressions (\ref{h68d}).
The figure also shows that $Q_{T_0}^e$ quickly increases with increasing atomic temperature $T_0$.
The relation $Q_{T_0}^e<Q_{T_0}^a$, obtained for $T_0<T$, 
indicates the dominance of heating of cold free atoms by the surface.
The substantial magnitude of the free-to-bound transition rate $G_{T_0}$ (dotted line)
indicates that a significant number of atoms can be adsorbed by the surface.
According to Fig. \ref{fig15}(b),
the free-to-free downward transition rate $Q_{T_0}^e$ (dashed line) crosses the 
upward transition rate $Q_{T_0}^a$ (solid line) when $T_0=T=300$ K, 
and then becomes the dominant decay rate. The relation $Q_{T_0}^e>Q_{T_0}^a$, 
obtained for $T_0>T$, indicates the dominance of cooling of hot free atoms by the surface.

\section{Conclusions}
\label{sec:summary}

In conclusion, we have studied the phonon-mediated transitions of an atom in a
surface-induced potential. We developed a general formalism, 
which is applicable for any surface--atom
potential. A systematic derivation of the corresponding density-matrix equation enables us to investigate the dynamics of both diagonal and off-diagonal elements.  
We included a large number of vibrational levels
originating from the deep silica--cesium potential.
We calculated the transition and decay rates from both bound and free levels.
We found that the rates of phonon-mediated transitions between translational levels depend on
the mean phonon number, the phonon mode density, and the matrix element of the force
from the surface upon the atom. Due to the effects of these competing factors,
the transition rates usually first increase and then reduce with increasing transition frequency. We focused on the transitions from bound states.
Two specific examples, namely, when the initial level is a shallow level also
when it can be one of the deep levels have been worked out.
We have shown that there can be marked differences in the absorption
and emission behavior in the two cases. 
For example, both the absorption and emission rates from the deep bound levels can be several
orders (in our case, six orders) of magnitude larger than the corresponding rates
from the shallow bound levels. We also analyzed various types of transitions from free states.
We have shown that, for thermal atomic cesium with temperature 
in the range from 100 $\mu$K 
to 400 $\mu$K in the vicinity of a silica surface with temperature of 300 K, 
the adsorption (free-to-bound decay) rate is about two times larger than the  heating (free-to-free upward decay) rate, while the cooling (free-to-free downward decay) rate is negligible.

\begin{acknowledgments}
We thank M. Chevrollier for fruitful discussions.
This work was carried out under the 21st Century COE program on ``Coherent Optical Science.''
\end{acknowledgments}


\begin{thebibliography}{99}

\bibitem[$*$]{a} Also at Institute of Physics and Electronics, Vietnamese Academy of Science and Technology, Hanoi, Vietnam.

\bibitem{our traps} 
V. I. Balykin, K. Hakuta, Fam Le Kien, J. Q. Liang, and  M. Morinaga,  
Phys. Rev. A \textbf{70}, 011401(R) (2004); 
Fam Le Kien, V. I. Balykin, and K. Hakuta, Phys. Rev. A \textbf{70}, 063403 (2004).

\bibitem{cesium decay} Fam Le Kien, S. Dutta Gupta, V. I. Balykin, and K. Hakuta, 
Phys. Rev. A \textbf{72}, 032509 (2005).

\bibitem{two atoms} Fam Le Kien, S. Dutta Gupta, K. P. Nayak, and K. Hakuta, 
Phys. Rev. A \textbf{72}, 063815 (2005). 

\bibitem{Lima} E. G. Lima, M. Chevrollier, O. Di Lorenzo, P. C. Segundo, and M. Ori\'{a}, 
Phys. Rev. A \textbf{62}, 013410 (2000).

\bibitem{Oria2006}  T. Passerat de Silans, B. Farias,  M. Ori\'{a}, and  M. Chevrollier,
Appl. Phys. B \textbf{82}, 367 (2006).

\bibitem{boundspon} Fam Le Kien and K. Hakuta, Phys. Rev. A \textbf{75}, 013423 (2007).

\bibitem{spectrum} Fam Le Kien, S. Dutta Gupta, and K. Hakuta, e-print quant-ph/0610067.

\bibitem{Kali} K. P. Nayak, P. N. Melentiev, M. Morinaga, Fam Le Kien, V. I. Balykin,
and K. Hakuta, e-print quant-ph/0610136.

\bibitem{Henkel} C. Henkel and M. Wilkens, Europhys. Lett. \textbf{47}, 414 (1999).

\bibitem{Gortel} Z. W. Gortel, H. J. Kreuzer, and R. Teshima, Phys. Rev. B \textbf{22},
5655 (1980).

\bibitem{Hoinkes} H. Hoinkes, Rev. Mod. Phys. \textbf{52}, 933 (1980).

\bibitem{Bogolubov} N. N. Bogolubov, Commun. of JINR, E17-11822, Dubna (1978);
N. N. Bogolubov and N. N. Bogolubov Jr., Elementary Particles and Nuclei (USSR) \textbf{11},
245 (1980).

\bibitem{Zwanzig} R. Zwanzig,  \textit{Lectures in Theoretical Physics},  
eds. W. E. Brittin, B. W. Downs, and J. Downs 
(Interscience, New York, 1961) Vol. 3, p. 106; 
G. S. Agarwal,  \textit{Progress in Optics},  ed. E. Wolf
(North-Holland, Amsterdam, 1973) Vol. 11, p. 3;
L. Mandel and E. Wolf, \textit{Optical Coherence and Quantum Optics}
(Cambridge, New York, 1995) p. 880.

\bibitem{Javanainen} J. Javanainen and M. Mackie, Phys. Rev. A \textbf{58},
R789 (1998); M. Mackie and J. Javanainen, \textit{ibid.} \textbf{60},
3174 (1999).

\bibitem{Luc-Koenig} E. Luc-Koenig, M. Vatasescu, and F. Masnou-Seeuws,
Eur. Phys. J. D \textbf{31}, 239 (2004).

\bibitem{Agrawal} See, for example, G. P. Agrawal, \textit{Nonlinear Fiber Optics}
(Academic, New York, 2001). 

\bibitem{coolingbook} H. J. Metcalf and P. van der Straten, \textit{Laser Cooling and Trapping} (Springer, New York, 1999).

\end{thebibliography}
\end{document}